\documentclass[9pt,conference]{IEEEtran}
\IEEEoverridecommandlockouts
\usepackage{cite}
\usepackage{amsmath,amssymb,amsfonts}
\usepackage{algorithmic}
\usepackage{graphicx}
\usepackage{makecell}
\usepackage{textcomp}
\usepackage{multirow}
\usepackage{multicol}
\usepackage{listings}
\usepackage{tikz}
\usepackage{xcolor}
\usepackage{subfigure}
\usepackage[all]{background}
\usepackage{stackengine}
\setstackEOL{\\}
\setstackgap{L}{\normalbaselineskip}
\SetBgContents{\color{gray}{\tiny \Longstack{PREPRINT - To appear at ASP-DAC 2024}}}% Set contents
\SetBgPosition{4.5cm,1cm}% Select location
\SetBgOpacity{1.0}% Select opacity
\SetBgAngle{0}% Select rotation of logo
\SetBgScale{1.8}% Select scale factor of logo

\usepackage{enumitem}
\setlist[itemize]{align=parleft,left=0pt..1em}
\colorlet{punct}{red!60!black}
\definecolor{background}{HTML}{EEEEEE}
\definecolor{delim}{RGB}{20,105,176}
\colorlet{numb}{magenta!60!black}

\newcommand*\circled[1]{\tikz[baseline=(char.base)]{
            \node[shape=circle,fill,inner sep=1pt] (char) {\textcolor{white}{#1}};}}

\lstdefinelanguage{json}{
    basicstyle=\ttfamily\scriptsize,
    numberstyle=\tiny,
    numbersep=8pt,
    showstringspaces=false,
    breaklines=true,
    frame=lines,
    backgroundcolor=\color{background},
    literate=
     *{0}{{{\color{numb}0}}}{1}
      {1}{{{\color{numb}1}}}{1}
      {2}{{{\color{numb}2}}}{1}
      {3}{{{\color{numb}3}}}{1}
      {4}{{{\color{numb}4}}}{1}
      {5}{{{\color{numb}5}}}{1}
      {6}{{{\color{numb}6}}}{1}
      {7}{{{\color{numb}7}}}{1}
      {8}{{{\color{numb}8}}}{1}
      {9}{{{\color{numb}9}}}{1}
      {:}{{{\color{punct}{:}}}}{1}
      {,}{{{\color{punct}{,}}}}{1}
      {\{}{{{\color{delim}{\{}}}}{1}
      {\}}{{{\color{delim}{\}}}}}{1}
      {[}{{{\color{delim}{[}}}}{1}
      {]}{{{\color{delim}{]}}}}{1},
}

\def\BibTeX{{\rm B\kern-.05em{\sc i\kern-.025em b}\kern-.08em
    T\kern-.1667em\lower.7ex\hbox{E}\kern-.125emX}}
\begin{document}
\thispagestyle{plain}
\pagestyle{plain}
\bstctlcite{IEEEexample:BSTcontrol}

\title{MemSPICE: Automated Simulation and Energy Estimation Framework for MAGIC-Based Logic-in-Memory\vspace{-0.4cm}}

\author{\IEEEauthorblockN{Simranjeet Singh\IEEEauthorrefmark{1}\IEEEauthorrefmark{5}, Chandan Kumar Jha\IEEEauthorrefmark{3}, Ankit Bende\IEEEauthorrefmark{5}. Vikas Rana\IEEEauthorrefmark{5}, \\ Sachin Patkar\IEEEauthorrefmark{1}, Rolf Drechsler\IEEEauthorrefmark{3}\IEEEauthorrefmark{4},
Farhad Merchant\IEEEauthorrefmark{2} \IEEEauthorblockA{\IEEEauthorrefmark{1}Indian Institute of Technology Bombay, India, \IEEEauthorrefmark{3}University of Bremen, Germany, \\ \IEEEauthorrefmark{5}Forschungszentrum Jülich GmbH, Germany, \IEEEauthorrefmark{4}DFKI GmbH, Germany,  \IEEEauthorrefmark{2}Newcastle University, UK}}
\{simranjeet, patkar\}@ee.iitb.ac.in, \{si.singh, a.bende, v.rana\}@fz-juelich.de, \\ \{chajha, drechsler\}@uni-bremen.de, farhad.merchant@newcastle.ac.uk \vspace{-0.5cm}}  

\maketitle
\begingroup\renewcommand\thefootnote{\textsection}
% \footnotetext{Equal contribution}
\endgroup

\begin{abstract}
%Digital Logic-in-Memory (LiM) has gained popularity as it allows the minimization of data movements.
% To the best of our knowledge, no existing framework is capable of generating a spice netlist from a hardware description language specifically for logic-in-memory (LiM) using memristors. 

Existing logic-in-memory (LiM) research is limited to generating mappings and micro-operations. In this paper, we present~\emph{MemSPICE}, a novel framework that addresses this gap by automatically generating both the netlist and testbench needed to evaluate the LiM on a memristive crossbar. MemSPICE goes beyond conventional approaches by providing energy estimation scripts to calculate the precise energy consumption of the testbench at the SPICE level. 
% Unlike current energy estimation techniques relying on coarse-grain methods for mappings obtained using MAGIC-based designs,~\textit{MemSPICE} recognizes the limitations of such approaches. 
We propose an automated framework that utilizes the mapping obtained from the SIMPLER tool to perform accurate energy estimation through SPICE simulations. To the best of our knowledge, no existing framework is capable of generating a SPICE netlist from a hardware description language. 
% The introduction of MemSPICE aims to alleviate concerns regarding energy consumption in IMC and provide more reliable and precise energy estimations. 
By offering a comprehensive solution for SPICE-based netlist generation, testbench creation, and accurate energy estimation, MemSPICE empowers researchers and engineers working on memristor-based LiM to enhance their understanding and optimization of energy usage in these systems. Finally, we tested the circuits from the ISCAS'85 benchmark on MemSPICE and conducted a detailed energy analysis.
% Additionally, our framework includes energy estimation scripts to calculate the energy consumption of the testbench at the SPICE level. Current energy estimation techniques for mappings obtained using MAGIC-based designs rely on coarse-grain methods, which we have found to significantly underestimate the energy consumption of MAGIC operations performed with a memristor crossbar. To address this issue, we propose an automated framework that utilizes the mapping obtained from the SIMPLER tool to perform accurate energy estimation through SPICE simulations. By doing so, we aim to alleviate concerns regarding energy consumption in IMC systems and provide more precise energy estimations.
\end{abstract}

\section{Introduction}
\label{intro}
In-memory computing using memristors is gaining popularity as it helps overcome the von Neumann bottleneck in traditional computing. Memristors possess two states: high resistive state (HRS) and low resistive state (LRS), which are mapped to Boolean logic `0' and `1' for logic-in-memory (LiM) implementation. Among various techniques like IMPLY~\cite{kvatinsky2013memristor}, FELIX~\cite{gupta2018felix}, and majority logic~\cite{DDH+:2023}, memristor-aided logic (MAGIC)~\cite{KBL+:2014} is widely adopted for LiM due to its superior energy and latency performance~\cite{EHA+:2021}. 
% \begin{figure}
%     \centering
%     \includegraphics[width=\linewidth]{Figures/NOR_gate_isolation.png}
%     \caption{MAGIC NOR implementation in a single row with isolation voltage}
%     \label{fig:NOR_iso}
% \end{figure}

\begin{figure}
    \centering
    \includegraphics[width=0.9\linewidth]{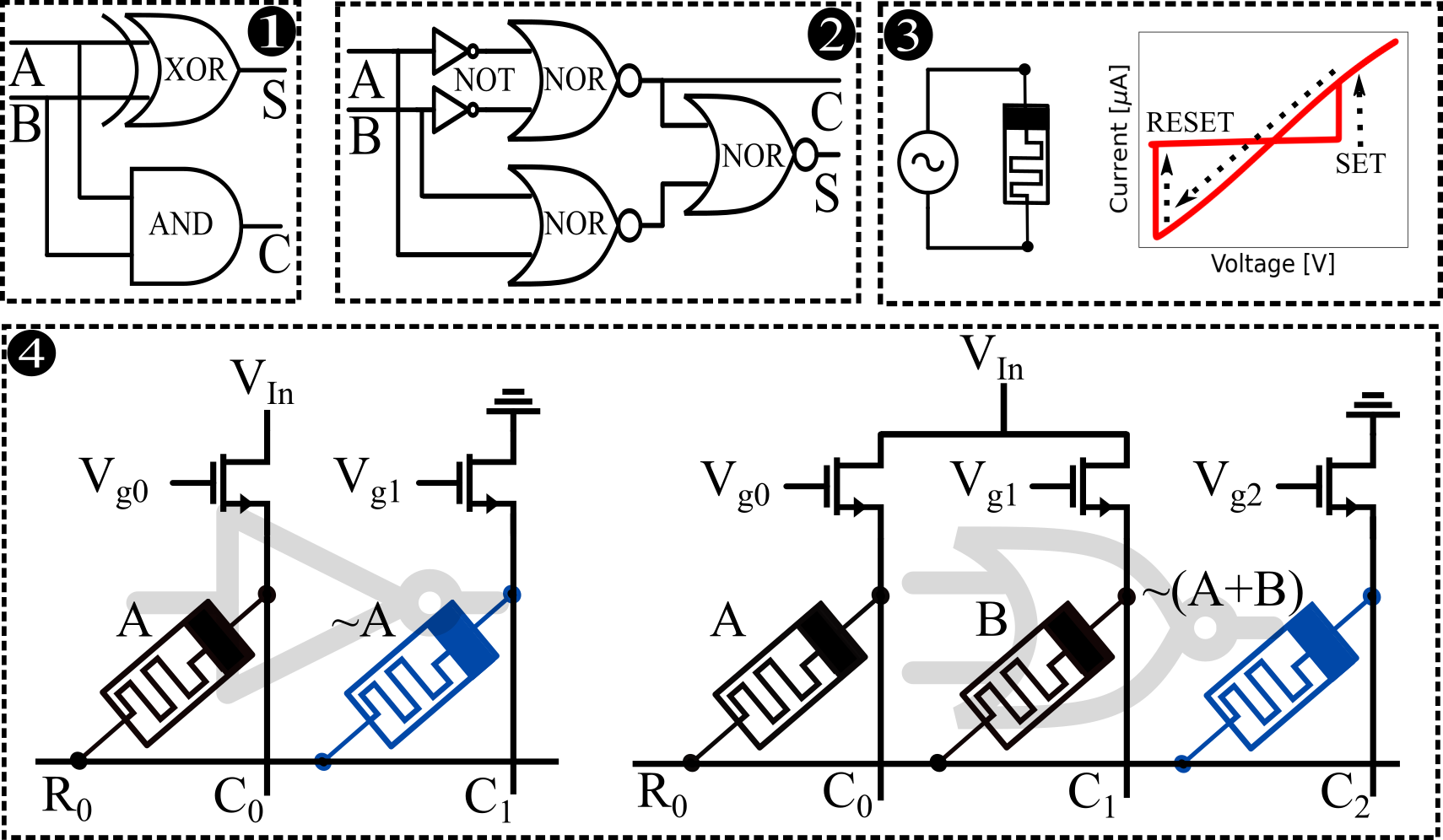}
    \caption{Mapping process of standard logic gates defined in hardware description language to MAGIC NOT and NOR gates using memristors. The figure showcases the I-V curve of the memristors, providing insights into their resistive states and electrical characteristics.} 
    % \color{red}{IV curve needs to be changed with VTEAM model IV} }
    \label{nornot}
\end{figure}

The larger design is synthesized to MAGIC NOR and NOT gates for a single-row memristor crossbar using the SIMPLER tool~\cite{RRA:2020}. The tool maps MAGIC operations to three memristors (two inputs, one output) and two memristors (one input, one output) on the crossbar. Additionally, it generates the necessary number of cycles and input/output memristors for a given application or benchmark. As an illustration, consider a netlist for a half adder provided in Fig.~\ref{nornot}~\circled{1}. Initially, it is synthesized into a NOR and NOT netlist, as depicted in Fig.~\ref{nornot}~\circled{2}. Subsequently, the NOR and NOT gates are sequentially mapped onto memristors connected in series. The resulting implementation of MAGIC NOR and NOT using memristors is presented in Fig.~\ref{nornot}~\circled{$4$}.

Given the increasing popularity of the MAGIC design style in mainstream computing~\cite{EHA+:2021}, evaluating this technique's energy consumption on larger circuit datasets is crucial. However, current methods for energy calculation rely on a coarse-grained approach, multiplying the average energy consumption of an operation by the number of occurrences in an application~\cite{thangkhiew2018efficient}. This approach is unsuitable for accurately estimating energy consumption in the MAGIC design style. A more detailed analysis of the implemented circuit is necessary to provide fine-grained energy values as presented~in~\cite{singh2023optimize}.

% \begin{figure}[!t]
%     \centering
%     \subfigure[]{\includegraphics[width=0.5\linewidth]{Figures/NOR_gate.png}}
%     \subfigure[]{\includegraphics[width=0.35\linewidth]{Figures/NOT.png}}
%     \caption{MAGIC design style based (a) NOR gate, and (b) NOT gate}
%     \label{nornot}
% \end{figure}

This paper proposes ~\emph{MemSPICE}, an automated SPICE-level simulation and accurate energy estimation framework for the MAGIC design style. The proposed framework automatically generates a SPICE-level netlist and testbench voltages for a given application/benchmark. Furthermore, it provides fine-grained energy numbers by calculating the energy consumed by each device in the crossbar, irrespective of its contribution to the operation. The framework empowers researchers to obtain accurate energy estimates for digital designs, offering valuable insights into their methodologies at the circuit level. Furthermore, circuit designers can utilize this framework to implement additional optimizations at the circuit level, showcasing their benefits on the entire crossbar rather than just individual gates. The followings are the contributions of this~paper:

% explain this in detail
\begin{itemize}
    \item Introducing MemSPICE, a framework that takes SIMPLER tool mapping (\textit{.json}) as input and autonomously generates the SPICE-level netlist. 
    % It allows the facility to change the device model and variations if necessary. 
    % Moreover, MemSPICE generates the voltage sequence for a given mapping for benchmarking.
    \item Detailed energy estimation techniques for MAGIC-based digital LiM at the low level, providing valuable insights to designers.
    \item Finally, the SPICE-level netlist for widely-used benchmarks (ISCAS'85) is automatically generated using MemSPICE, and a comparative analysis of the energy consumption values with current methodologies is conducted0.
\end{itemize}
% --In-Memory computing, emphasis on MAGIC \\
% -- Energy techniques \\
% -- Emphasis on the energy consumption due to isolation voltages \\
% -- Motivation for our proposed work \\

The rest of the paper is organized as follows. Section~\ref{backnrw} discusses the necessary background and related work. In Section~\ref{sec:Methodology}, we discuss the MemSPICE methodology in detail. In  Section~\ref{sec:Experimental_results}, we discuss the benchmark circuit preparation, the results obtained using MemSPICE methodology, and compare the results with state-of-the-art methods. We conclude the paper in Section~\ref{sec:conclusions}. 

% The prior works have limited the energy comparisons to single NOR and NOT operations. They multiply this with the number of operations to report the coarse gain energy numbers. This method has severe limitations as was also shown in one of the works [cite ESWEEK]. 

% \textcolor{red}{We want to enable researchers to get an accurate estimate of their contributions at the synthesis level, mapping strategies, etc. to gain insights of their methodologies at the circuit level. On top of that the circuit designers can also use this framework to build on top optimizations at the circuit level to show their benefits on the entire crossbar rather than individual gates.}
% {\color{red}{change the intro to fit atleast one page}}

\section{Background and Related Work}
\label{backnrw}
\subsection{Memristive Devices or Memristors}
 Memristive devices or memristors are two-terminal passive devices with variable resistance. When a voltage is applied across the terminals of a memristor, the resistance changes in response to the magnitude and direction of the current flow. Fig.~\ref{nornot}~\circled{3} depicts the electrical characterization of a memristor, showcasing its I-V curve with distinct SET and RESET points marked. Importantly, even when the power is turned off, the memristor retains its resistance value until a new voltage is applied, making it a non-volatile memory element~\cite{Sebastian2020nature}. Due to their unique characteristics, memristors have garnered significant interest in various fields, such as in-memory computing, neuromorphic computing, and LiM applications. The maximum and minimum resistance values of memristors are represented by LRS and HRS, which are mapped to logic states `0' and `1', respectively. Considering these states, Boolean logic operations can be performed by arranging the memristive devices into crossbar connections and applying different voltages across them. Numerous models have been proposed in the literature to characterize the memristive devices for SPICE simulation. VTEAM model~\cite{KRE+:2015} is one of the models that can characterize various memristive technologies and is used for this work.

% Memristive devices or memristors are two-terminal passive devices with variable resistance, initially proposed by Chua in 1971~\cite{ch}. Memristors have non-volatile resistance states, which is a function of applied voltage amplitude and its duration, alternatively integral over time of the applied voltage across them. The maximum and minimum resistance values of memristors are represented by a low resistive state (LRS) and a high resistive condition (HRS), which are mapped to logic states '0' and '1', respectively. Considering these states, Boolean logic operations can be performed by arranging the memristive devices into crossbar connections and applying different voltages across them. Numerous models have been proposed in the literature to characterize the memristive devices for SPICE simulation. VTEAM model~\cite() is one of the models that can characterize a variety of different memristive technology and hence being used for this work. 
\subsection{MAGIC Design Style}

MAGIC is a stateful logic technique that utilizes memristive devices to implement logic operations, where the resistive states of memristors store the inputs and outputs of these operations. The MAGIC design style incorporates NOR and NOT gate implementations. To perform MAGIC operations, an initialization step is required, initializing the output memristor to logic `1' before executing the operations. During a NOR operation, an execution voltage ($V_{in}$) is applied to the input memristors ($A$, $B$), while the output memristor is connected to the ground, as illustrated in Fig.~\ref{nornot}~\circled{4}. In contrast, the NOT operation only requires two memristors. Since the NOR gate serves as a universal gate, any logic function can be achieved by combining these gates sequentially. The initial step involves converting the given Verilog design netlist into MAGIC NOR and NOT netlists. Fig.~\ref{nornot} showcases the mapping of the netlist given in~\circled{1} to~\circled{2}, which is done using the SIMPLE mapping tool.

% Figure 2(b) shows the implementation of MAGIC NOR operation in a 1x6 crossbar. In a single time sequence or operation cycle, only three memristors are active (two for inputs, marked in black, and one for output, marked in green). The other three inactive memristors (marked in red) need to be isolated from the network for the functional correctness of the operation. The isolation voltage also solves the purpose of undesired switching of inactive devices. However, applying isolation voltages consumes energy per operation, which needs to be considered for accurate energy estimation of an application. 

\subsection{SIMPLER Mapping Tool}
\begin{figure}
    \centering
    \includegraphics[width=0.9\linewidth]{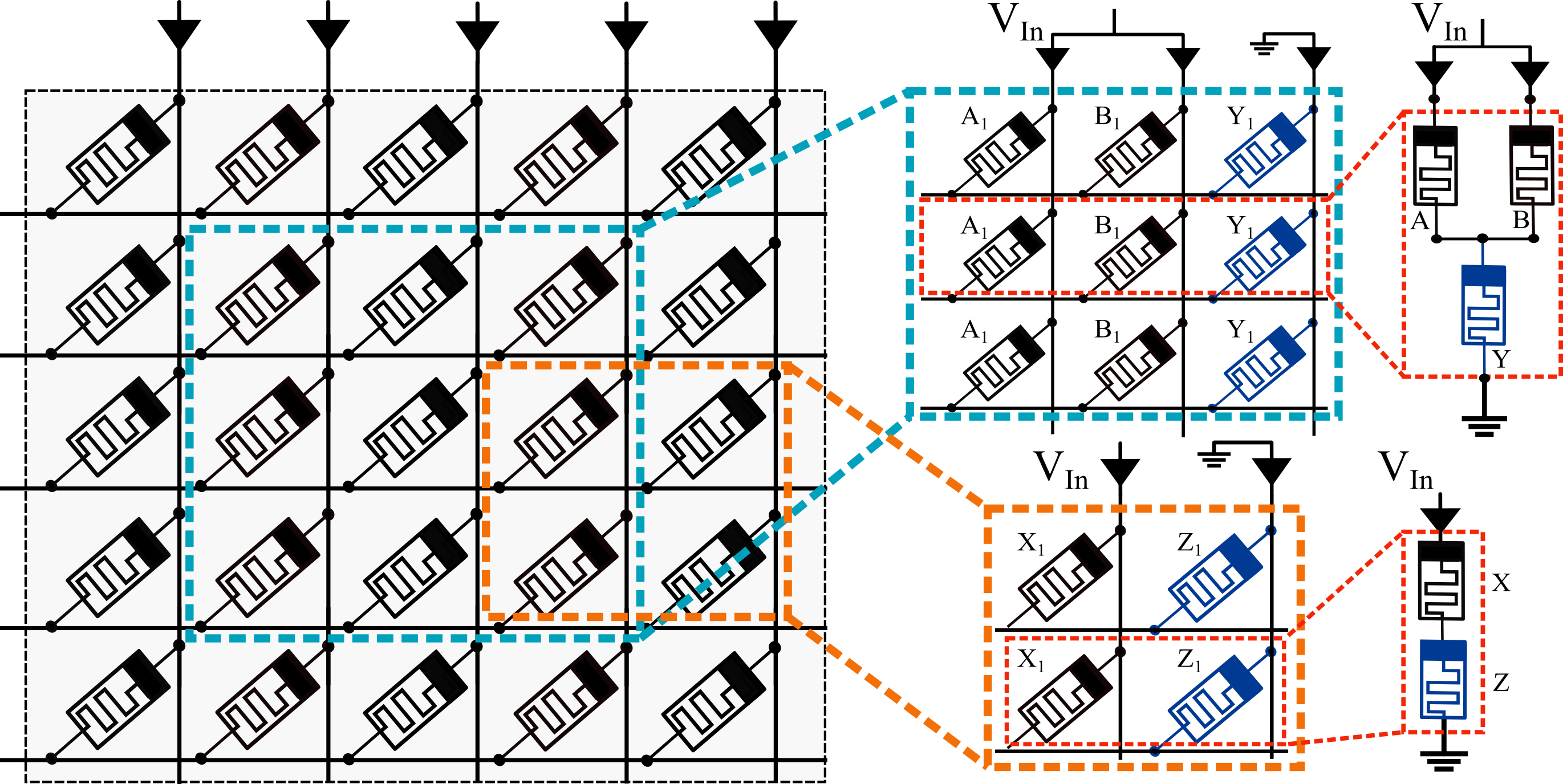}
    \caption{Demonstrating the simultaneous execution of MAGIC NOT and NOT gates in parallel SIMD fashion. The row-wise parallel approach depicted in the Figure.}
    \label{fig:simd}
\end{figure}

SIMPLER MAGIC is a synthesis and mapping tool that is used to generate the MAGIC design style-based mapping of any arbitrary design~\cite{RRA:2020}. The SIMPLER tool generates the mapping of the design to one row of a memristor crossbar.
The work shows that these designs can then be replicated across multiple rows of the crossbar, which can operate on different data. Another advantage of doing the same is that the controller and additional circuits cost can be amortized as they can be shared across the multiple rows of the crossbar as they are performing the same set of operations. Hence the mapping obtained using SIMPLER is inherently suitable for ~\emph{single instruction multiple data} (SIMD) instructions as shown in Fig.~\ref{fig:simd}. Since it supports SIMD-based operations it is useful for applications that require higher throughput. It also supports reusing the same memristors cells to map a larger design to a crossbar of limited size.

% SIMPLER is a synthesis tool to map MAGIC design style based operations to a single row of memristors~\cite{RRA:2020}. It uses graph theory and compiler register allocation technology to optimize the execution sequences. The number of memristors on which the design needs to be mapped can be selected. Given a logic function defined in a hardware description language, the SIMPLER tool automatically generates a sequence of MAGIC design style-based NOR and NOT operations for a single memory row. The same mapping can be replicated across multiple rows to enhance the throughput~\cite{RRA:2020} for~\emph{single instruction multiple data} (SIMD) instructions. The NOR gate implementation inside the crossbar is shown as an example. All the rows can similarly perform the same NOR operation on different data. It allows reusing the cells to save the area or exploiting the parallelism of MAGIC to improve latency whenever needed. 

% \subsection{SIMD}
% {\color{red}{write about SIMD - Chandan}}

\subsection{Related Work}
The automated flow to generate for analog implementation has been presented in the literature~\cite{antoniadis2022opensource}. The automation is achieved through Cadence skill programming, making it suitable for specific applications. Additionally, the work in~\cite{memorycompiler2021} automates the attachment of peripherals to the RRAM memory, though limited to memory functionality. Various simulators have been proposed at different levels of abstraction, including system, architecture, circuit, and device levels~\cite{Staudigl:2022}. Some approaches have also explored mixed-signal simulation, integrating different levels together~\cite{Fritscher:2019}. However, these existing approaches are often limited to a single device type and fixed parameters. In contrast, our paper introduces a digital LiM implementation, expanding the usability of RRAM beyond conventional memory and analog applications. The proposed implementation allows for configurable parameters, providing flexibility in simulations and enabling more extensive exploration of LiM designs.

Moving toward energy consumption techniques, the current methods for calculating energy consumption involve multiplying the \textit{average energy used during an operation by the number of such operations} in an application, which is a highly coarse-grained approach to determine the energy consumed by the MAGIC design style~\cite{thangkhiew2018efficient}. Surprisingly, despite its popularity, this methodology falls short of providing accurate estimates of the energy dissipated by an application since it does not account for the energy consumed during initialization, reading, and loading input patterns. As far as we know, no existing framework has the capability to conduct an in-depth analysis of energy consumption in LiM. This paper addresses this gap by introducing the MemSPICE framework, which enables detailed energy consumption analysis for LiM designs by running a SPICE-level simulation.
\section{MemSPICE FRAMEWORK}
\label{sec:Methodology}

\begin{figure*}
    \centering
    \includegraphics[width=0.9\linewidth]{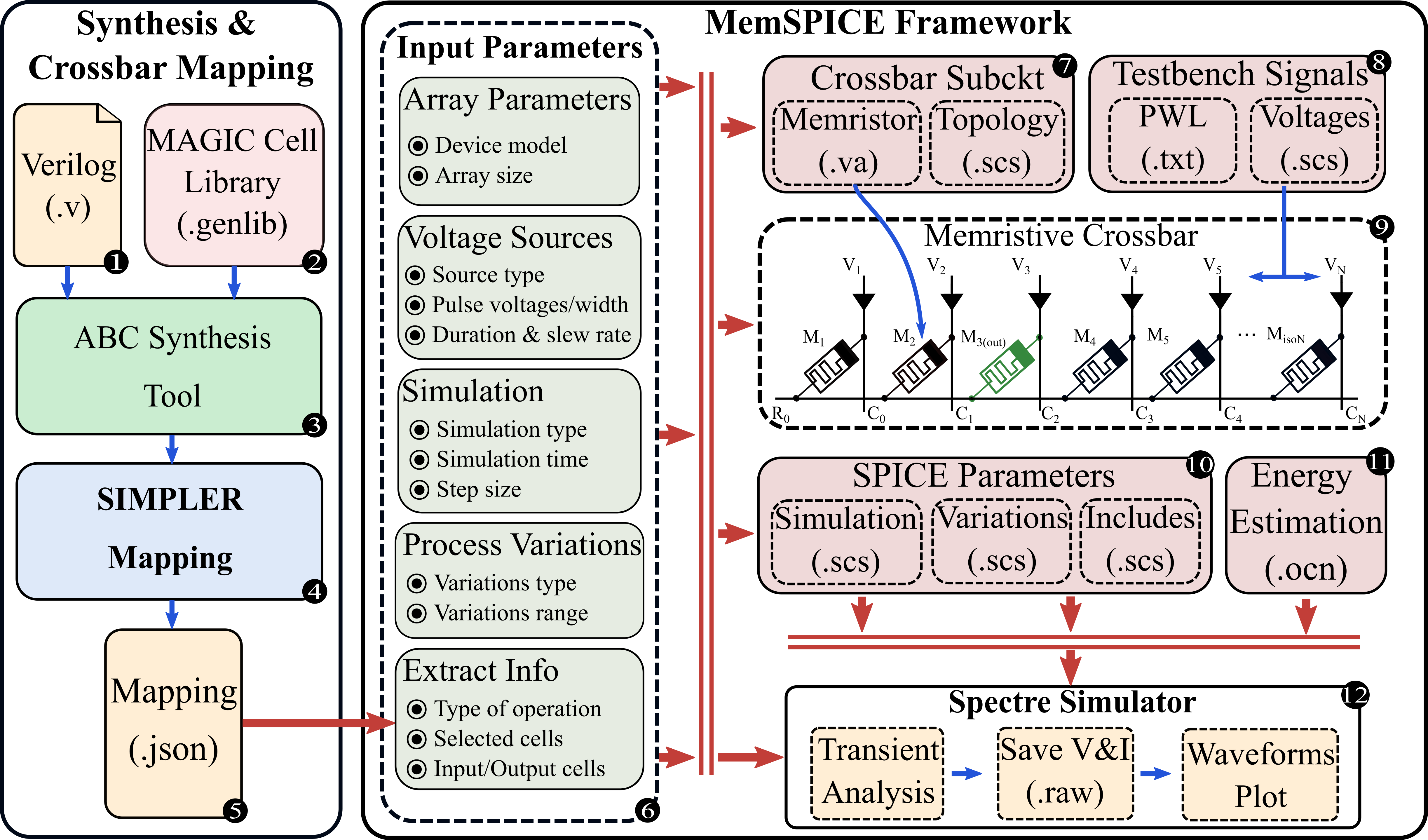}
    \caption{MemSPICE, an automated digital synthesis framework for LiM computing. It has two major blocks, the first to synthesize the Verilog design on possible logic gated on the crossbar, and the second block automatically generates the SPICE-level simulation files.\vspace{-0.5cm}}
    \label{fig:method}
\end{figure*}

This section presents the MemSPICE framework to obtain accurate energy estimates by mapping the Verilog design to the MAGIC design style at the SPICE level. The MemSPICE methodology is shown in Fig~\ref{fig:method}. It is an automated framework to generate the SPICE-level netlist for MAGIC NOR and NOT gate. The automated process comprises three-step (A) SIMPLER tool mapping, (B) automated SPICE-level netlist, (C) MemSPICE output \& test bench creation, and (D) energy estimations. In the following section, we discuss each step in detail.
% This section discusses the overall MemSPICE framework to perform the automated netlist generation, testbench generation, and energy estimation. The overall MemSPICE framework is shown in Fig~\ref{fig:methdod}. We will now discuss the framework in detail.

\subsection{Verilog Synthesis}
The process starts with the Verilog design synthesis using the ABC synthesis tool~\cite{mishchenko2007abc}. As an illustration, we consider a half-adder example, and you can find the Verilog declaration in Listing~\ref{lst:half_adder_verilog}. This Verilog file (Fig.~\ref{fig:method}~\circled{1}) serves as an input, alongside the MAGIC cell library (NOT and NOR gates) shown in Fig.~\ref{fig:method}~\circled{2}, to the ABC tool. Subsequently, the ABC tool synthesizes (Fig.~\ref{fig:method}~\circled{3}) the Verilog description into NOR and NOT gates. The resulting NOR/NOT netlist becomes the input for the SIMPLER mapping tool (Fig.~\ref{fig:method}~\circled{4}), which then generates an optimal mapping for MAGIC design style gates. Additionally, the SIMPLER mapping tool performs a sequential mapping of the MAGIC NOT and NOR operations, utilizing three and two memristors on the crossbar, respectively. Crucial information such as the required number of cycles, input/output memristors, and other relevant details tailored to the specific application or benchmark is generated by the SIMPLER tool, and this data is stored in a .json file (Fig.~\ref{fig:method}~\circled{5}). Listing~\ref{lst:half_adder} shows the .json file representing the mapping of the half-adder on five memristors connected in a single row. Further, this .json file is used to generate the SPICE-level netlist and test vectors for simulation and detailed energy analysis.

\subsection{Automated SPICE Netlist}
MemSPICE utilizes the output from the SIMPLER tool in .json format as input to automatically generate the SPICE-level netlist. From this \textit{.json} file, MemSPICE extracts critical information such as input and output data devices and the sequence of NOR/NOT operations for execution. Additionally, MemSPICE takes several parameters as input to customize the simulation process, which is presented in Fig.~\ref{fig:method}~\circled{6}. These parameters are organized into four categories based on their role during the simulation:
\begin{enumerate}[label=\roman*)]
    \item Array parameters: These encompass device model and array size arguments essential for the simulation. As the SIMPLER tool maps the design to a single row, the generated SPICE-level netlist also contains a single row with `n' numbers of columns.
    \item Voltage-related parameters: MemSPICE supports pulse and piece-wise linear (PWL) voltage sources and allows control over parameters such as pulse amplitude, width, period, and rise/fall time of the voltage source for testbench purposes.
    \item Simulation-related parameters: These parameters offer flexibility in choosing the type of simulation (DC simulation, transient) required for testing, along with defining simulation time and step size during the simulation.
    \item Process variations: This set of parameters controls the type and range of variation for the devices. When process variation parameters are set, MemSPICE draws variation values from a normal distribution probability function around the mean value, with a defined standard deviation in input parameters.
\end{enumerate}
By effectively managing these input parameters, MemSPICE empowers users to tailor the SPICE-level netlist generation and simulation according to their specific requirements, enhancing the flexibility and accuracy of the overall analysis. Considering all the parameters, MemSPICE formulates a spectre-compatible netlist (.scs) in a crossbar configuration. Subsequently, this crossbar SPICE netlist is assembled into a crossbar symbol, complete with dedicated inputs and outputs optimized for benchmarking purposes. In the output, MemSPICE generates multiple .scs files, encompassing the crossbar sub-circuit, testbench signals, as well as simulation and energy estimation files, all essential for conducting SPICE simulations.

% -- (rough) explain in detail, for each device, we take two terminals and add device parameters before exporting to .scs file. For benchmarking, we need to attach voltage sources to each terminal of the device and connect them in a crossbar structure. The tool automatically connects them in a crossbar structure and attaches a voltage source to the common terminal (r0 -rn). 
% To generate the automated SPICE-level netlist, MemSPICE considers the output of the SIMPLER tool (.json) as an input and extracts the critical information such as input and output data devices and execution sequences of NOR/NOT operations. It also considers the device model (verilog.a model or path of std cell) to map the operation on the device model. It also provides the features to consider the variations of the devices during execution sequences, such as device-to-device variation and cycle-to-cycle. 

\begin{lstlisting}[language=verilog,numberstyle=\tiny,basicstyle=\ttfamily\scriptsize,caption={Verilog logic of half adder as an input to MemSPICE.},captionpos=b,label={lst:half_adder_verilog},frame=single,linewidth=0.98\linewidth]
module half_adder (A, B, S, Cout);
//Input Ports Declarations
input A, B;
//Output Ports Declarations
output S, Cy;
//Logic
assign S = A ^ B ;
assign Cy = (A & B);
endmodule
\end{lstlisting}
% --adding for little space.
\begin{lstlisting}[language=json,numberstyle=\tiny,basicstyle=\ttfamily\scriptsize,caption={Mapping of a half adder onto five memristors in a row.},captionpos=b,label={lst:half_adder},frame=single,linewidth=0.98\linewidth]
"Row size": 5,
"Number of Gates": 5,
"Inputs": "{A(0),B(1)}",
"Outputs": "{S(4),Cy(2)}",
"Reuse cycles": 1,
"Execution sequence": {
"T0": "Init{'D(2)','D(3)','D(4)'}",
"T1": "n5_(4)=inv1{A(0)}",
"T2": "n6_(3)=inv1{B(1)}",
"T3": "Cy(2)=nor2{n6_(3),n5_(4)}",
"T4": "Init{n5_(4),n6_(3)}",
"T5": "n8_(3)=nor2{B(1),A(0)}",
"T6": "S(4)=nor2{n8_(3),Cy(2)}"}

\end{lstlisting}

\subsection{MemSPICE Output \& Test-benching }
% -- explain all the files generated by MemSPICE with code snippets.
MemSPICE generates various files in the output, all compatible with spectre simulation. These files are individually created and called in a single main file to execute the final simulation. 

%\begin{enumerate}[label=\roman*)]
\subsubsection{Crossbar Subckt} MemSPICE first considers a device model (.va) and creates a symbol based on it. Crossbar Subckt block as shown in Fig.~\ref{fig:method}~\circled{7} connects symbols in crossbar topology as defined by the input arguments. To adhere to the MAGIC design style's requirement of switches at each row and column, MemSPICE automatically attaches switches at the crossbar pins for convenient access. An example of the crossbar sub-circuit file for the implementation of a half-adder is depicted in Listing~\ref{lst:crossbar}. In this listing, the crossbar comprises six pins ($r_7$ and $c_0$-$c5$), with a total of six devices connected between these pins. Additionally, seven switches are connected to memristor pin a .scs file containing this information is exported as shown in Listing~\ref{lst:peripherals}.
\begin{lstlisting}[language=,numberstyle=\tiny,basicstyle=\ttfamily\scriptsize,caption={Automated crossbar sub circuit for half adder implementation.},captionpos=b,label={lst:crossbar},frame=single,linewidth=0.98\linewidth]
\\ Connect memirstors together in crossbar
subckt crossbar_sub r0   c0 c1 c2 c3 c4 c5 
I0 (r0 c0 n0)  VTEAM_model <model parameters> 
I1 (r0 c1 n1)  VTEAM_model <model parameters> 
---
ends crossbar_sub
\\ call sub circuit for pheripherls
crossbar0 (sub0_r0  sub0_c0 sub0_c1 sub0_c2 
sub0_c3 sub0_c4 sub0_c5) crossbar_sub
\end{lstlisting}

\begin{lstlisting}[language=,numberstyle=\tiny,basicstyle=\ttfamily\scriptsize,caption={Switch connected to rows and columns of the crossbar subckt.},captionpos=b,label={lst:peripherals},frame=single,linewidth=0.98\linewidth]
\\ Relay for modelling transistor as switch
W0 (0 sub0_r0 v_r0 0) relay ropen=1T rclosed=1.0 
W1 (v_c0 sub0_c0 v_s0 0) relay ropen=1T rclosed=1.0 
W2 ---
---
\end{lstlisting}

\subsubsection{Testbench Signals} 
The testbench signals encompass voltage sources and their timing values, which are dependent on the execution sequence and the type of operation within that sequence. These voltages are crucial for performing operations and initializing or loading input data to the devices. Additionally, the selection of the device for each operation is determined by the voltages on switches, extracted from the execution sequence within the .json file. As the operations on the devices are highly flexible, Fig.~\ref{fig:method}~\circled{8} takes all these variations into consideration and generates voltage sources for the crossbar pins along with the PWL file as shown in Listing~\ref{lst:volt}. The PWL file contains voltage waveforms in text format, essential for ensuring the functional correctness of the operations.
\begin{lstlisting}[language=,numberstyle=\tiny,basicstyle=\ttfamily\scriptsize,caption={Voltage source connected to rows and columns of the \\ crossbar for operation, the actual value of voltage amplitude and \\ duration are defined in PWL files.},captionpos=b,label={lst:volt},frame=single,linewidth=0.98\linewidth]
\\ Create voltage source with PWL file path
V0 (v_r0 0) vsource type=pwl file = "./pwl/r0.txt" 
V1 (v_c0 0) vsource type=pwl file = "./pwl/c0.txt" 
---
V7 (v_s0 0) vsource type=pwl file = "./pwl/s0.txt" 
---
\end{lstlisting}

To achieve digital LiM using memristors, five different voltage sources are required, each tailored to specific aspects of the operation and ensuring the proper functioning of MAGIC operation. 
\begin{itemize}
    \item \textbf{\textit{Input Voltage:}} The input voltage determines the resistances of the memristors used as inputs. It is mapped to 2.0V for storing `1' (LRS) and 0.0V for HRS (memristors' default state).

    \item \textbf{\textit{Read Voltage:}} The read voltage is applied to read the state of the output memristors after all operations. It is set at 0.2V, relatively low compared to SET and RESET voltages, ensuring reliable read operations. There is potential for reducing energy consumption by further decreasing the read voltage.

    \item \textbf{\textit{Initialization Voltage:}} The intermediate output memristors, storing intermediate results, are initialized to LRS for accurate operation. This voltage configures them to LRS and is mapped to 2.0V for correct operation.

    \item \textbf{\textit{Switch Voltage:}} During the execution cycle, only selected devices are active, isolated from others using a switch voltage. The switch voltage of the chosen devices is set to 2.0V (ON), while others remain at 0V (OFF), preserving their state during computation.

    \item \textbf{\textit{Operation Voltage:}} During operation, specific voltages are applied to memristors for NOR and NOT operations in MAGIC design style. Input memristors are connected to 1.0V during NOT operation, while the output memristor is connected to the ground.
\end{itemize}

%\item
\subsubsection{SPICE Simulation Parameters} Fig.~\ref{fig:method}~\circled{10} is responsible for managing the parameters essential for the simulation tool. These parameters encompass the path of the model and other peripherals within the circuit. Additionally, if there are any variations to be considered during the simulation, this block handles the variation values which are mapped to the devices during simulation. It also maps the input pattern to the corresponding device state. Moreover, it includes simulation analysis parameters that dictate how the simulation will be conducted. MemSPICE automatically calculates the simulation time based on the total execution cycles and the duration of pulses for each operation. However, other parameters, such as step size and include path, are analyzed from the input parameters and then parsed into the simulation format. 

As depicted in Listing~\ref{lst:SPICE_param}, this step involves including the path of the memristor model and specifying parameters like input pattern and variations, if applicable. Additionally, the current values for each memristor are saved during the simulation, and these values are later utilized for energy analysis. By following this process, the parameters are effectively integrated and employed during the simulation, thereby enhancing the accuracy and efficiency of the overall analysis.

%\end{enumerate}

\begin{lstlisting}[language=,numberstyle=\tiny,basicstyle=\ttfamily\scriptsize,caption={SPICE simulation parameters.}, captionpos=b,label={lst:SPICE_param},frame=single,linewidth=0.98\linewidth]
\\ include memristor verilog model
ahdl_include "<path_to>/verilog.va"
parameters in0=0 in1=2 <input & variation parameters>
\\ save the current for futher analysis
save crossbar0.I0:n ---
tran tran stop=2.4e-08 step=1e-12 maxstep=1e-12 
\end{lstlisting}

% \begin{figure*}

% \begin{multicols}{5} % Specify the number of columns
% \lstset{language=Python} % Set the language for the listings

% \begin{lstlisting}[caption={$V_{c0}$},frame=single,linewidth=0.98\linewidth]
% 0 0
% 1e-12 in0
% 1.301e-09 in0
% 1.302e-09 0
% 3e-09 0
% --

% \end{lstlisting}

% \columnbreak % Move to the next column

% \begin{lstlisting}[caption={$V_{c1}$},captionpos=b,label={lst:peripherals},frame=single,linewidth=0.98\linewidth]
% 0 0
% 1e-12 in1
% 1.301e-09 in1
% 1.302e-09 0
% 3e-09 0
% ---

% \end{lstlisting}
% \columnbreak % Move to the next column

% \begin{lstlisting}[caption={$V_{c2}$},captionpos=b,label={lst:peripherals},frame=single,linewidth=0.98\linewidth]
% 0 0
% 1e-12 2
% 1.301e-09 2
% 1.302e-09 0
% 3e-09 0
% --

% \end{lstlisting}

% \columnbreak % Move to the next column

% \begin{lstlisting}[caption={$V_{r0}$},captionpos=b,label={lst:peripherals},frame=single,linewidth=0.98\linewidth]
% 0 0
% 1e-12 2
% 1.301e-09 2
% 1.302e-09 0
% 3e-09 0
% ---

% \end{lstlisting}

% \columnbreak % Move to the next column

% \begin{lstlisting}[caption={$V_{s0}$},captionpos=b,label={lst:peripherals},frame=single,linewidth=0.98\linewidth]
% 0 0
% 1e-12 2
% 1.301e-09 2
% 1.302e-09 0
% 3e-09 0
% ---

% \end{lstlisting}

% \end{multicols}
% \end{figure*}

\begin{figure}
    \centering
    \includegraphics[width=0.9\linewidth]{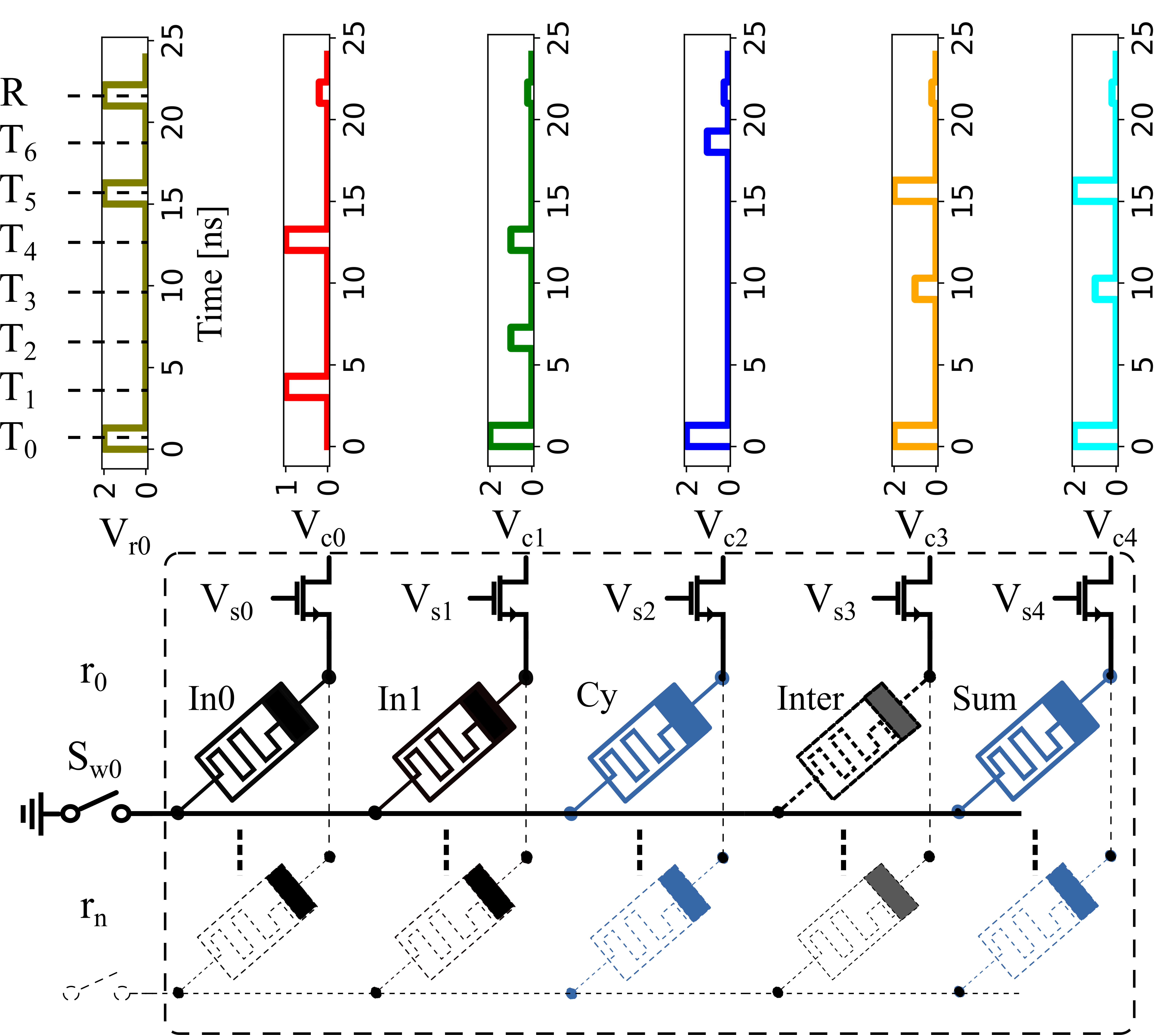}
    \caption{Half adder implementation for ``10" input ($in_0=1$, $in_1=0$) using five memristors in a row as per the mapping in Listing~\ref{lst:half_adder}. This implementation requires a total of 7 execution cycles, including an initialization ($T_0$) and a reused cycle ($T_4$). Similar to a SIMD architecture, the same operation can be executed concurrently in different rows ($r_n$). \vspace{-0.3cm} }
    \label{fig:half_adder_wave}
\end{figure}

\subsection{Energy Estimation}
To accurately calculate energy consumption, we simulate the circuit (Fig.~\ref{fig:method}~\circled{12}) and generate a waveform file. Energy per device is calculated using Equation~\ref{equ:eng}, summing the product of voltage and current over simulation time. 
\begin{equation}
\label{equ:eng}
 % \resizebox{0.3\textwidth}{!}{
\rm Energy =\sum_{i=0}^{n} \int_{0}^{t} (V_i \times I_i) dt
% }
\end{equation}
`n' denotes memristor count, and `t' represents simulation time based on pulse width and cycles required to complete the benchmark. The equation covers all memristor activity, including initialization, execution, and read energy.

% To accurately calculate the energy consumption, we perform a simulation as given in Fig.~\ref{fig:method}~\circled{12} and generate a waveform file. The energy is calculated regardless of the device's selection. The energy consumed by single device is determined using the Equation~\ref{equ:eng}, which sums up the product of voltage and current over simulation time.
% \begin{equation}
% \label{equ:eng}
%  % \resizebox{0.3\textwidth}{!}{
% \rm Energy =\sum_{i=0}^{n} \int_{0}^{t} (V_i \times I_i) dt
% % }
% \end{equation}
% Here, `n' represents the number of memristors, and `t' corresponds to the simulation time. The simulation time depends on the pulse width and the total number of cycles necessary to complete the benchmark. The given equation captures the activity on each memristor irrespective of its use in the cycle, which includes initialization, execution, and read energy.
The total energy consumed for a given application is computed by combining Equation~\ref{equ:eng} for each device within the network. These equations are stored as a .ocn file, as shown in Fig.~\ref{fig:method}~\circled{11}. This file serves as a repository for the energy equations, facilitating their systematic application and enabling the determination of the overall energy consumption of the network.

Fig.~\ref{fig:half_adder_wave} presents the schematic and waveform generated for the half-adder implementation, which is synthesized from Listing~\ref{lst:half_adder_verilog}. Notably, the half adder implementation is mapped to a single row of 5 memristors. As shown in Listing~\ref{lst:half_adder}, it also requires memristor reinitialization to store intermediate results. These intermediate results are represented by the dotted line with memristors labeled as `Inter', while the final sum and carry are stored in the memristors labeled as `Sum' and `Cy', respectively.

The implementation necessitates six voltage sources ($V_{r0}$ and $V_{c0} -V_{c4}$) with values tailored to the logic implementation during each execution sequence. Additionally, there are five other voltage sources responsible for opening switches to execute operations. Furthermore, these operations can be performed in SIMD fashion, but they are beyond the scope of this work.

In summary, the framework offers digital and circuit designers the ability to thoroughly test and optimize their designs in a more accurate and automated manner, providing valuable insights into the effectiveness of their methodologies.

% \textcolor{red}{Show how the voltage file looks like }
% -- How we are generating PWL file (testbench) \\
% -- Ocean script to calculate the energy \\
% \textcolor{red}{Show the ocean script template}
% -- energy of each cell in every cycle. \\
% \textcolor{red}{Show the current and voltage waveform. Also, show the integral equation used in cadence}

\section{Experimental Results}
\label{sec:Experimental_results}

\begin{table*}[!t]
    \caption{Energy Consumption Results on ISCAS'85 Benchmarks}
    \label{tab:benchmark}
    \setlength{\tabcolsep}{3pt}
    \centering
    \begin{tabular}{|c|c|c|c|c|c|c|c|c|c|c|c|c|c|c|c|c|c|}
    \hline
          Circuit & PI/PO & Cycles & NOT & NOR&Reinit& \makecell{Energy (pJ) } &\multicolumn{3}{c|}{ \textbf{I1:} Energy (pJ), Input; all 0} &\multicolumn{3}{c|}{\textbf{I2:} Energy (pJ), Input; all 1} & \multicolumn{3}{c|}{\textbf{I3:} Energy (pJ), Input; alt.}\\
         \cline{8-16}
        & & & & & &Literature~\cite{jha2022,thangkhiew2018efficient} &  Exe & Init & Total & Exe & Init   & Total & Exe &  Init & Total\\
         \hline
         c17 & 5/2 & 14 &7 &6 & 0& 0.655&   0.674 &1161 & 1161.67 &  0.90 & 1164 &1164.9 &  0.75 &1162 & 1162.75\\
        c432 & 36/7 &250 &101 &148& 0&12.31  & 15.23&1142& 1157.23 & 15.96&1163& 1178.96& 14.92&1153 & 1169.92 \\
        c499 & 41/32 & 605 &213 &390& 1&29.6& 30.85&1790 &1820.85 & 25.53&1803 &1828.53& 28.45 &1788 &1816.45\\
          c880 & 60/26 & 505 &194 &310 &1& 24.85& 22.91 & 1819 & 1841.91& 25.26& 1785 & 1810.26& 24.76 & 1779 & 1803.76\\
          c1908 & 33/25 & 571 &210 &359&1& 27.99 &26.33 &1850 &1876.33& 22.95 &1814 & 1836.95& 25.72 & 1822 & 1847.72 \\
          c3540 & 50/22 & 1397 &465 &928 &3& 68.18&37.92 & 2676 &2713.92&37.51 & 2474 &2511.51&38.22 & 2431 &2469.22 \\
        \hline
        % \multirow{11}{*}{\textbf{ISCAS'89}} & c27 & 7/4 & 12 &5 &7 & 0.5944&  127 & 126.9 & 126.39\\
        %  & c208 & 19/10 & 100 &45 &55 & 4.977&  142.77 & 144.25 & 143.315\\
        % & c344 & 24/26 & 156 &60 &96 & 7.691&  154.61 & 155.74 & 155.239\\
        % & c386 & 13/13 & 185 &62 &123 & 9.055&  157.205 & 157.205 & 159.92\\
        % & c420 & 35/18 & 198 &87 &111 & 9.84&  163.93 & 164.5 & 162.68\\
        % & c510 & 25/13 & 312 &115 &197 & 15.34&  187.115 & 182.18 & 184.409\\
        % & c820 & 23/25 & 367 &136 &231 & 18.05&  198.622 & 191.002 & 194.431\\
        % & c832 & 23/24 & 347 &120 &227 & 17.01&  195.60 & 187.473 & 191.10\\
        % & c838 & 67/34 & 389 &166 &223 & 19.29&  204.16 & 202.23 & 203.59\\
        % & c1488 & 14/25 & 722 &207 &514 & 35.04&  1133.08 & 1108.63 & 1118.88\\
        % & c1494 & 14/25 & 737 &208 &528 & 35.74&  1133.08 & 1116.11 & 1125.31\\
        % \hline
    \end{tabular}
    
\end{table*}

This section presents the results achieved through the MemSPICE framework. To validate the methodology, we conducted tests using the ISCAS'85 benchmark with the proposed framework. To ensure comparability with existing literature~\cite{singh2023optimize}, we mapped all benchmarks to a single row of 512 devices. For operating, programming, and reading the state of memristors, we employed different voltage with a 1.3 ns pulse width, with 1 ps rise and fall times.

Table~\ref{tab:benchmark} presents ISCAS'85 benchmark results. It includes benchmark names, primary inputs/outputs count, cycles needed for the final result, NOR/NOT gate counts, re-initialization cycle, and energy values from the literature. MemSPICE allows testing with input patterns for energy consumption. I1, I2, and I3 represent all 0's, all 1's, and alternating 1's and 0's, respectively. We now discuss the results in detail using some examples from the benchmarks. 

% The result for ISCAS'85 benchmark  are shown in Table~\ref{tab:benchmark}. The first column has the name of the benchmark suite and their respective benchmark circuits. The PI/PO column gives the number of primary inputs and primary outputs in respective circuit. The 'Cycles' column gives the number of cycles required to obtain the final result for the given benchmark circuit. The 'NOR' and the 'NOT' columns give the number of the NOR and NOT required for implementation, respectively. The next column gives the energy number provided in the literature. MemSPICE allows to test the circuit with different input patterns. In the last three columns, we show the energy consumption of the various benchmark circuit using three different input patterns, I1, I2, and I3 respectively. The pattern P1 denotes all 0's at the input, the pattern I2 denotes all 1's at the input, and the pattern PI denotes alternating 1's and 0's at the input. We now discuss the results in detail using a couple of examples from the benchmarks. 

\subsection{c432 from ISCAS 85} 
We mapped the c432 benchmark circuit (27-channel interrupt controller) on 512 devices, featuring 36 inputs and 7 outputs. Its operation requires 250 cycles, with 101 NOT operations and 148 NOR operations. The c432 is relatively a small benchmark that easily fits within 512 devices without any reinitialization cycles. The execution energy closely matches the literature, but the initialization energy is considerably higher. As a first step, MemSPICE initializes all devices to `1' (LRS) except for input memristors, which are necessary for MAGIC implementation. In this case, initializing 476 devices (512-PI) to `1' dominates the initialization energy. The execution energy varies significantly based on the input pattern due to the benchmark's smaller size. This is true for all the benchmark circuits that can be mapped to far fewer memristors than 512. The unused memristor energy dominates the operation energy in this case.

\begin{figure*}[!t]
    \centering
    \subfigure[]{\includegraphics[width=0.45\linewidth]{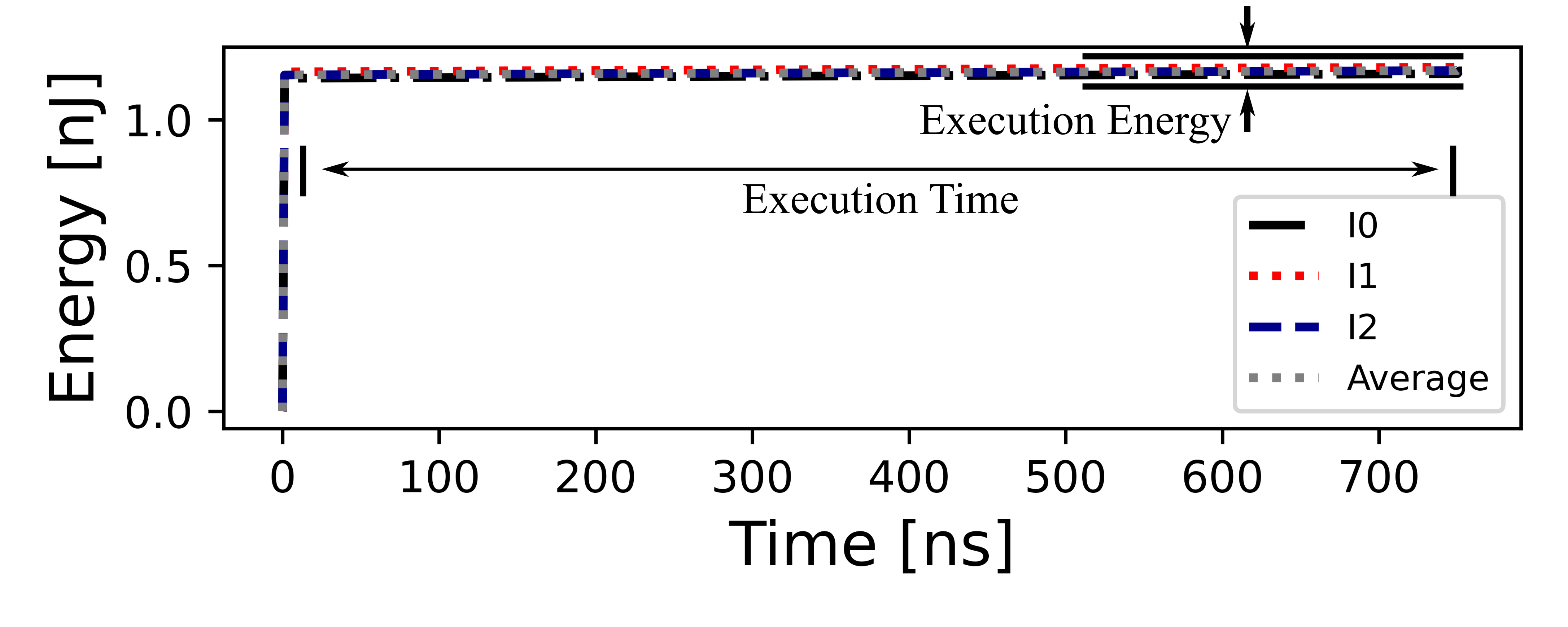}\label{c432_a}} \vspace{-.1cm}
    \subfigure[]{\includegraphics[width=0.25\linewidth]{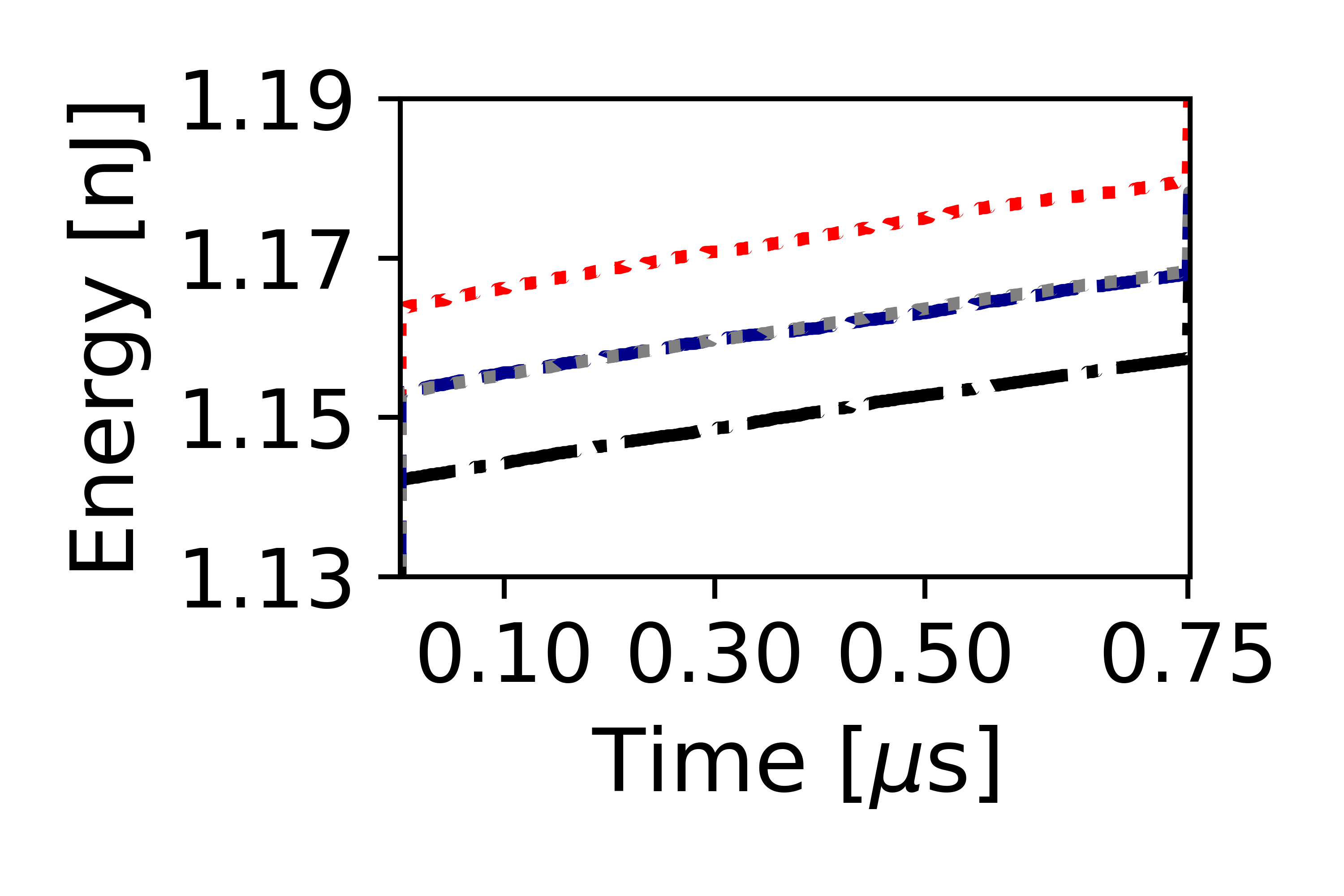}\label{c432_b}}
    \subfigure[]{\includegraphics[width=0.25\linewidth]{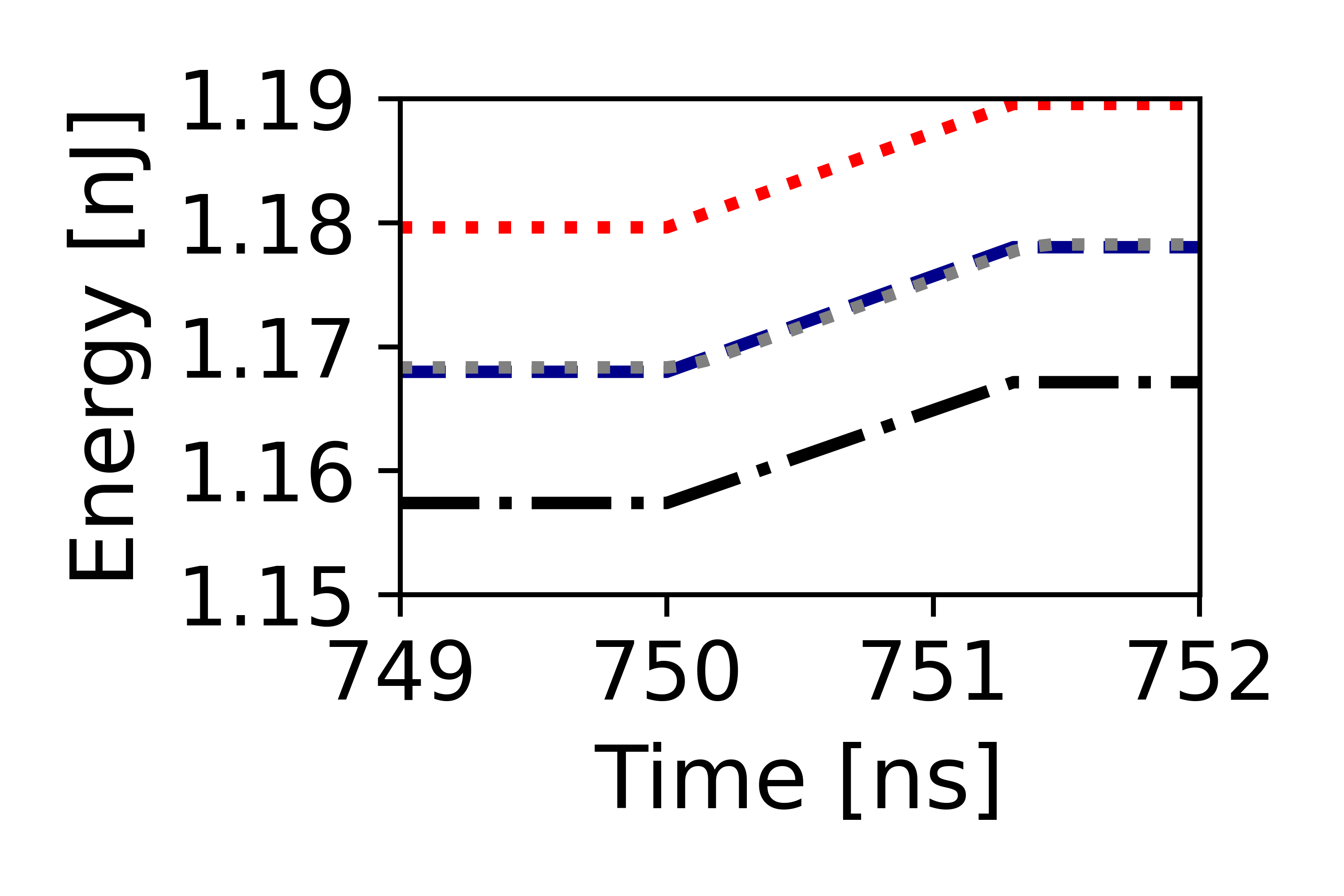}\label{c432_c}}
    \vspace{-0.7\baselineskip}
    \caption{Energy consumption of c432 benchmark (a) total energy, (b) execution energy and (b) read energy.}
    \vspace{-.2cm}
    \label{fig:c432}
\end{figure*}

Fig.~\ref{fig:c432} presents the energy breakdown of the c432 benchmark. Since c432 does not have any re-initialization cycle, all the devices are initialized simultaneously in the first cycle for use in the operation cycle. A closer view of Fig.~\ref{c432_a} is provided in~\ref{c432_b} and~\ref{c432_c} to illustrate the execution and read energy, respectively. The read energy is computed by reading the state of each memristor in a row to verify functionality and switching during operation.

% We see that the c17 benchmark circuit has 5 inputs and 2 outputs. The number of cycles required for the operation is 14. The circuit consists of 7 NOT gates and 6 NOR gates. The current methods that only evaluate the energy of the operation give 0.65 pJ as the overall energy consumption. We obtained that the energy consumption of the circuit is much higher when we evaluate using the MemSPICE framework. The obtained energy consumption for all three different patterns of inputs P1, P2, and P3 are 127 pJ. The difference in the energy consumption value is 195 $\times$ as compared to the current state of the art. This clearly highlights the need for the MemSPICE framework. The difference in energy consumption can be attributed to the inclusion of initialization energy and the energy consumption of all the other memristors that are not being used in a given cycle. We also see that for this benchmark irrespective of the input pattern the energy consumption is almost similar as the initialization energy required for the other unused memristors is much higher than the operation energy. This is true for all the benchmark circuits that can be mapped to far fewer memristors than 512. Since the crossbar size is fixed the unused memristor energy dominates the operation energy in this case. Overall, this further highlights the need for the MemSPICE framework. 

\subsection{c3540 from ISCAS 85}
As shown in Table~\ref{tab:benchmark}, the c3540 benchmark, which is an 8-bit arithmetic and logic unit,  requires 1397 cycles with 50 input and 22 output memristors. It comprises 465 NOT gates and 928 NOR gates. With the circuit mapped to 512 devices, three re-initialization of devices are needed to complete the entire operation, making re-initialization energy a significant factor. The energy consumption using the MemSPICE framework for I1, I2, and I3 input patterns is found to be 2676 pJ, 2474 pJ, and 2431 pJ, respectively. During re-initialization, we apply a SET voltage (2.0V) to reset the device to LRS state without reading its current state. If the device is already in the LRS state before re-initialization, it draws a considerable amount of current, significantly increasing the energy consumption. In this specific case, the re-initialization energy is approximately 70 $\times$ higher than the execution energy. 

\begin{figure*}[!t]
    \centering
    \subfigure[]{\includegraphics[width=0.4\linewidth]{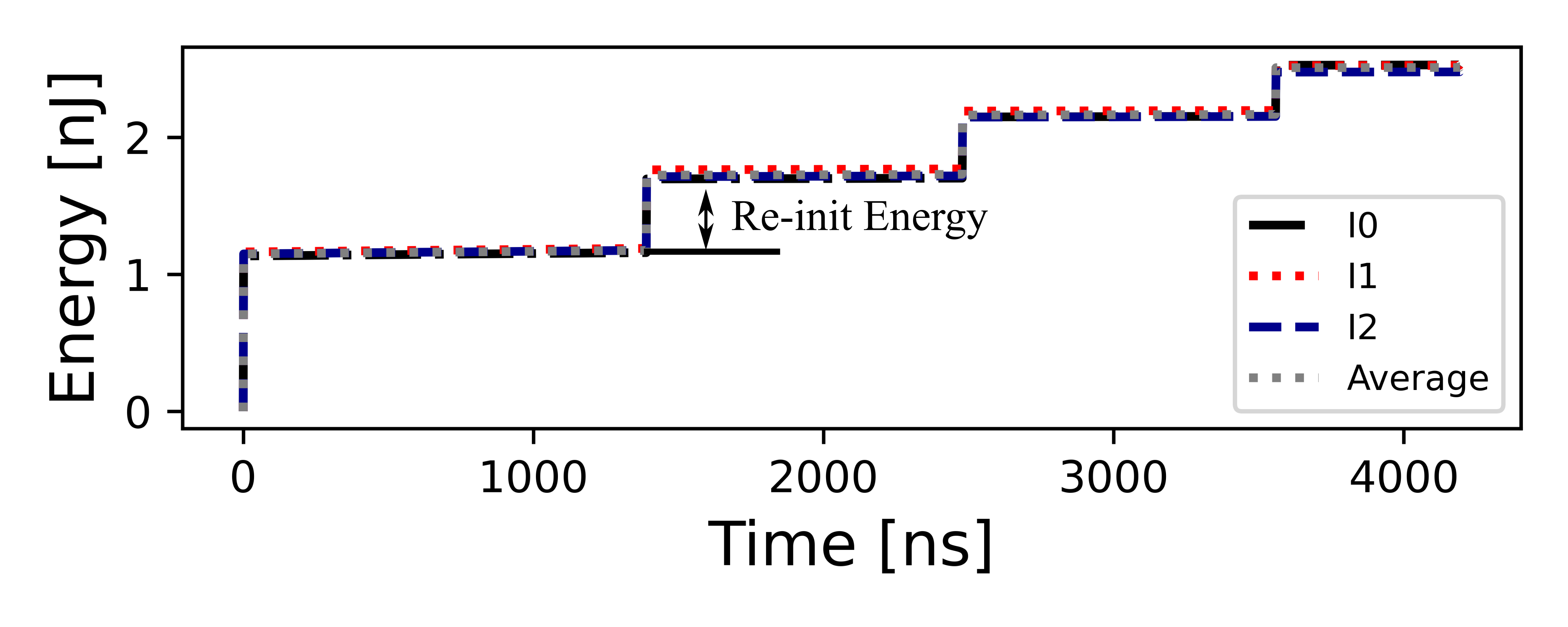} \label{c3540_a}}
    \subfigure[]{\includegraphics[width=0.25\linewidth]{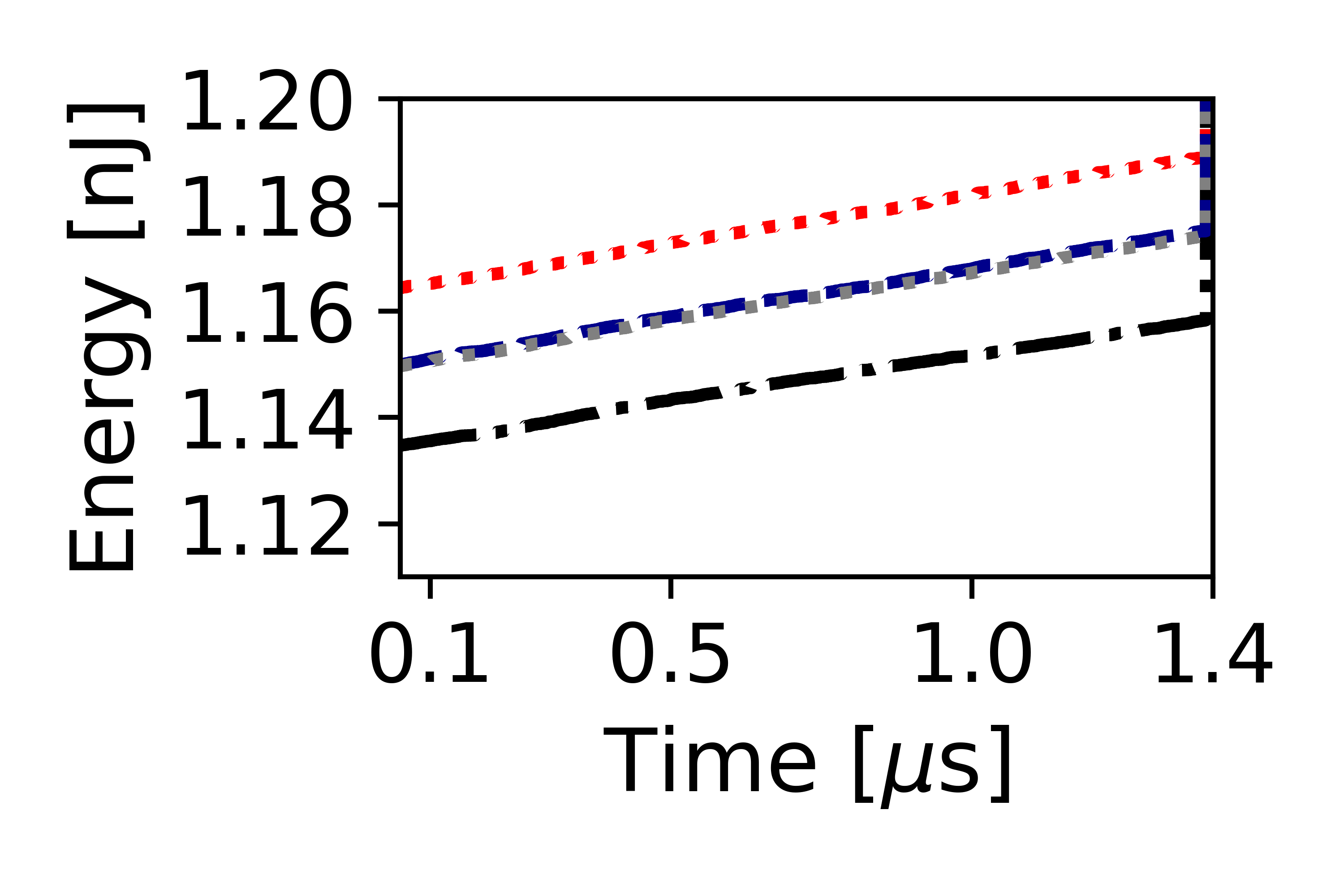} \label{c3540_b}}
    \subfigure[]{\includegraphics[width=0.25\linewidth]{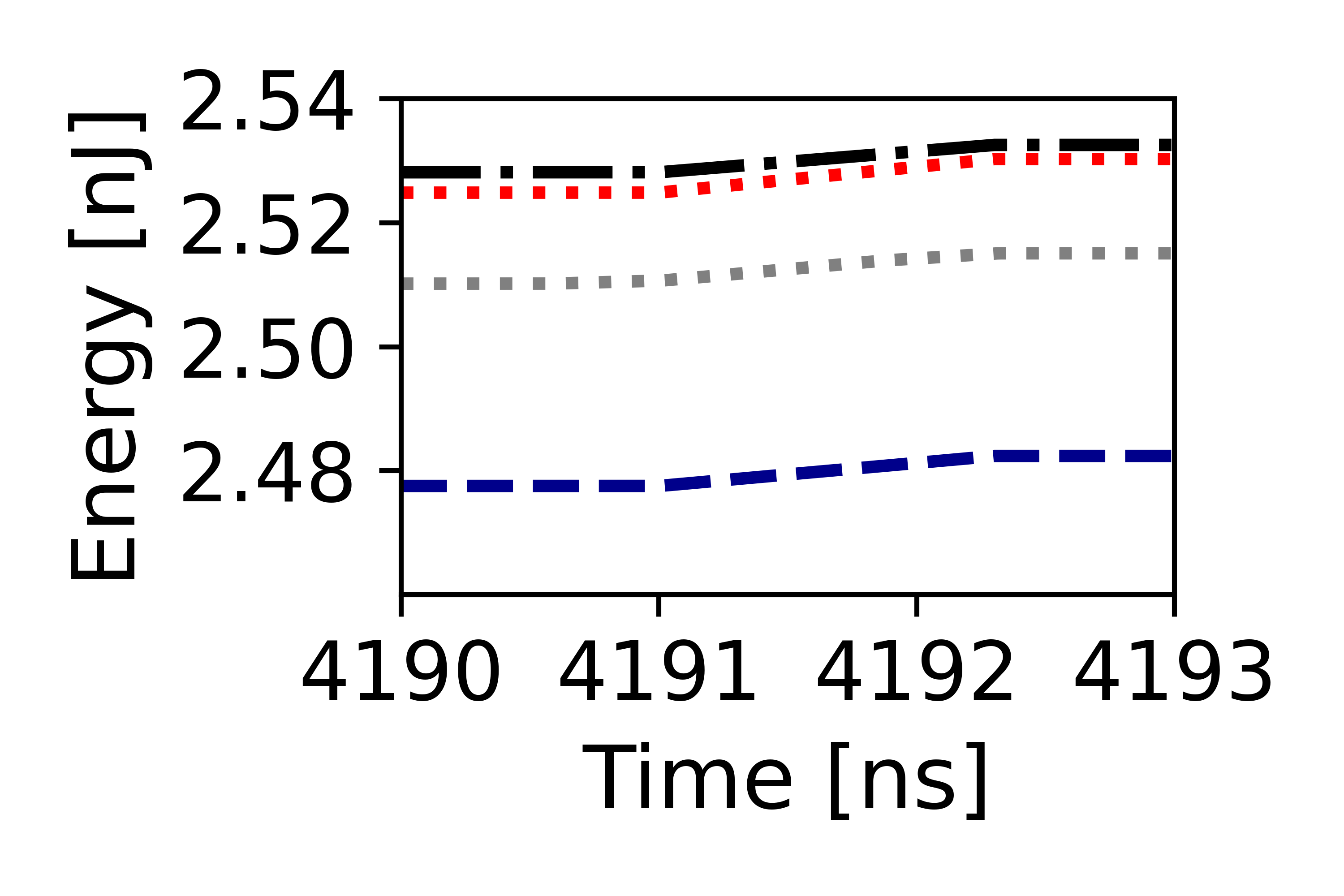}  \label{c3540_c}}
     \vspace{-0.8\baselineskip}
    \caption{Energy consumption of c3540 benchmark (a) total energy, (b) execution energy before first reintilization and (b) read energy.}
    \vspace{-0.4cm}
    
    \label{fig:c3540}
\end{figure*}

Fig.~\ref{fig:c3540} displays the energy breakdown of the c3540 benchmark. As indicated in Table~\ref{tab:benchmark}, c3540 requires three initialization cycles. The graph in Fig.~\ref{c3540_a} clearly illustrates the significant increase in energy during the re-initialization cycle. This graph highlights that the re-initialization energy dominates the energy consumption during execution. Fig.~\ref{c3540_b} and Fig.~\ref{c3540_c} provide a closer view of Fig.~\ref{c3540_a}.

% We see that the c3540 benchmark circuit has 50 inputs and 22 outputs. The number of cycles required for the operation is 1397. The circuit consists of 465 NOT gates and 928 NOR gates. The current methods that only evaluate the energy of the operation give 68.18 pJ as the overall energy consumption. We obtained the energy consumption using the MemSPICE framework for P1, P2, and P3 to be 
% 2920 pJ, 3015 pJ, and 2974 pJ respectively. The difference in the energy consumption value is 44 $\times$ as compared to the current state of the art. This can be attributed to the inclusion of initialization energy for the memristors and also during the operation, the entire crossbar consumes energy and not only the memristors used for the operation. The difference in energy consumption compared to the state-of-the-art method is lower than the c17 benchmark circuit as the design is larger than the entire crossbar and reuses the crossbar to perform the operations. The difference in the energy consumption between P1,P2 and P3 highlights that the operation energy dominates the initialization energy, and depending upon the input pattern different energy values will be obtained. This is another advantage of using the MemSPICE framework as it allows us to capture the energy consumption of the design depending on the input patterns. 

Table~\ref{tab:benchmark} presents the results for all the benchmarks, showcasing how the MemSPICE framework simplifies netlist generation and energy evaluation for MAGIC design style-based circuits. The execution energy values obtained by MemSPICE align closely with state-of-the-art energy calculations, validating the correctness of the methodology. Furthermore, MemSPICE can capture additional energy aspects not addressed in the existing literature. The utilization of SPICE-level simulation~\cite{singh2023optimize} in MemSPICE contributes to more precise energy values, emphasizing the framework's significance for accurate energy analysis and automation in digital LiM.

The output of the framework includes waveforms illustrating the executed operations in the sequence, with an explicit read pulse incorporated. To validate the functional correctness of the benchmarks, manual testing is performed, ensuring the accuracy of the files generated by the proposed framework. %However, for larger designs, a formal verification method becomes necessary to ensure comprehensive validation.

% {\color{red}{add two graphs for c432 and c3540 to show the result}}

% \begin{figure}
%     \centering
%     \includegraphics[width=\linewidth]{Figures/energy.png}
%     \caption{Caption}
%     \label{fig:enter-label}
% \end{figure}

\section{Conclusions}
\label{sec:conclusions}
% {\color{red}{rewrite the conclusion}}

This paper has introduced MemSPICE, a SPICE-level framework that fills the gap in existing research by automating the generation of SPICE netlists for digital LiM using memristors. MemSPICE offers users the ability to control and fine-tune various parameters, providing remarkable flexibility to accommodate diverse configurations and scenarios. It takes Verilog-defined logic and automatically generates SPICE-level simulations for detailed analysis. Notably, it provides precise and fine-grained energy values, enhancing the accuracy of energy estimation. The energy values generated by MemSPICE were found to be in agreement with the energy calculations reported in the literature. MemSPICE will enable researchers, without any circuit knowledge, to accurately estimate the benefits of their proposed ideas by performing automated SPICE-level simulations and we believe this will have a huge impact in this domain of research.
% In the future, we aim to provide an automatic formal verification methodology suitable for larger designs.
%believe that researchers can accurately estimate the impact of their  

\section*{Acknowledgments}
% \vspace{-1mm}
This work was supported in part by the Federal Ministry of Education and Research (BMBF, Germany) in the project NEUROTEC II under Project 16ME0398K, Project 16ME0399, German Research Foundation (DFG) within the Project PLiM (DR 287/35-1, DR 287/35-2) and through Dr. Suhas Pai Donation Fund at IIT~Bombay.

%\textcolor{red}{manually change the references to make it compact }
\bibliographystyle{IEEEtran}
\bibliography{Bib}

\end{document}